\documentclass[preprint,5p,times,twocolumn]{my-elsarticle}

\usepackage{tensor}
\usepackage{amssymb}
\usepackage{amsmath}
\usepackage{mathrsfs}

\biboptions{compress}

\usepackage[normalem]{ulem}
\usepackage[colorlinks=true,linkcolor=black,citecolor=blue,urlcolor=blue,filecolor=black]{hyperref}
\allowdisplaybreaks

\begin{document}

\begin{frontmatter}

\title{Boundary Freedom in Higher Spin Gravity}

\author[Urbana]{Arnaud Delfante}

\address[Urbana]{Department of Physics, University of Illinois, 1110 W. Green St., Urbana IL 61801-3080, U.S.A.\\
\href{mailto:delfante@illinois.edu}{delfante@illinois.edu}}

\begin{abstract}

We investigate new boundary conditions in three-dimensional asymptotically Anti–de Sitter gravity coupled to higher-spin fields, allowing for arbitrary boundary degrees of freedom. This generalization gives rise to an enlarged algebra of asymptotic charges and predicts novel anomalies in the dual field theory, whose physical origin is traced to corner ambiguities in the symplectic structure. We also prescribe a smooth flat limit, thereby extending these anomalous higher-spin effects to asymptotically flat spacetimes.

\end{abstract}

\end{frontmatter}

%% \linenumbers

%% main text 

%%%%%%%%%%%%%%%%%%%%%%%%%%%%%%%%%%%%%%

\section{Introduction} \label{sec. Intro}

In three dimensions, Fronsdal’s fields with spin $s>1$ do not carry local propagating degrees of freedom. This feature has made $D=3$ a privileged setting for the study of the spin two case~\cite{Staruszkiewicz:1963zza,Deser:1983tn,Deser:1983nh}. Despite the absence of radiation, three-dimensional gravity remains highly nontrivial for two reasons. First, the theory admits black hole solutions~\cite{Banados:1992wn}. Second, its asymptotic symmetry algebra exhibits an enhancement relative to the isometries of the vacuum~\cite{Brown:1986nw,Barnich:2001jy,Carlip:2005tz}. More precisely, the seminal work of Brown and Henneaux (BH)~\cite{Brown:1986nw} established that Anti-de Sitter (AdS$_3$) gravity with Dirichlet boundary conditions encompasses a centrally extended double copy of the Virasoro algebra, later identified as the mode algebra of the boundary stress tensor~\cite{Strominger:1997eq}. This result provided one of the earliest realizations of the AdS/CFT correspondence~\cite{Maldacena:1997re,Witten:1998qj}.

Beyond pure gravity, the simplified structure of three-dimensional models offers an attractive arena for exploring higher-spin extensions without the technical obstacles present in $D>3$. Already in~\cite{Aragone:1983sz} it was observed that in three dimensions consistent interactions do not require an infinite tower of higher-spin fields. One can consider, for instance, a finite multiplet containing all integer spins between $2$ and some maximal spin~$s$. This allows one to exploit the topological nature of the interactions and recast the theory as a Chern--Simons~(CS) gauge theory~\cite{Achucarro:1986uwr,Witten:1988hc,Blencowe:1988gj,Bergshoeff:1989ns,Vasiliev:1989re}. The spectrum studied in this work is governed by the gauge algebra $\mathfrak{sl}(s,\mathbb{R}) \oplus \mathfrak{sl}(s,\mathbb{R})$, with the AdS isometry algebra $\mathfrak{so}(2,2)$ principally embedded. Alternative choices of embeddings may yield different spectra \cite{Campoleoni:2024ced}.

The analysis of asymptotic symmetries in such higher-spin CS theories was initiated in~\cite{Henneaux:2010xg,Campoleoni:2010zq}, where it was shown that the algebra is given by two copies of the Zamolodchikov $\mathcal{W}_3$-algebra~\cite{Zamolodchikov:1985wn}, with central charge matching the BH value. This enhancement of the Einstein--Hilbert charge algebra suggested the possibility of higher-spin versions of the AdS/CFT correspondence and triggered an extensive body of research in this direction (see, e.g.,~\cite{Gaberdiel:2012uj}). In parallel, purely gravitational extensions of the BH analysis were developed by relaxing boundary conditions and gauge fixings~\cite{Compere:2013bya,Troessaert:2013fma,Perez:2016vqo,Grumiller:2016pqb,Campoleoni:2018ltl,Alessio:2020ioh,Geiller:2021vpg,Campoleoni:2022wmf,Ciambelli:2023ott,Delfante:2024npo}.

In this Letter, we initiate a bridge between these two directions. Using the framework of asymptotic symmetries, we extend the anomalous boundary conditions of three-dimensional CS gravity to the higher-spin case. This produces an enlarged charge algebra that does not preserve the on-shell action, thus predicting new anomalies in the dual theory. Following \cite{Ciambelli:2024vhy}, we also address the physical interpretation of these so-called ``kinematical'' charges by relating them to ambiguities in the symplectic structure, thereby motivating a distinction with the ``dynamical'' ones that generate the double copy of~$\mathcal{W}_3$. Finally, in analogy with the spin-$2$ case~\cite{Campoleoni:2018ltl,Ciambelli:2020eba,Campoleoni:2023fug}, we prescribe the behavior of boundary higher-spin fields for small values of the boundary speed of light, deducing a smooth flat limit of the bulk AdS analysis. This extends the flat-space higher-spin studies of~\cite{Afshar:2013vka,Gonzalez:2013oaa}, yielding a larger set of asymptotic symmetries as well as anomalous predictions for the putative boundary theory (see, for instance, \cite{Donnay:2023mrd}).

%%%%%%%%%%%%%%%%%%%%%%%%%%%%%%%%%%%%%%

\section{Boundary Conditions} \label{sec. BC}

We describe the coupling of massless higher-spin fields to $\text{AdS}_3$ gravity through the subtraction of two CS actions~\cite{Achucarro:1986uwr,Witten:1988hc,Blencowe:1988gj,Henneaux:2010xg,Campoleoni:2010zq}:
\begin{equation} \label{eq. CS_generic_split}
    S = S_{\mathrm{CS}}[\text{A}] - S_{\mathrm{CS}}[\bar{\text{A}}] \, ,
\end{equation}
with
\begin{equation} \label{eq. actionCS}
    S_{\mathrm{CS}}[\text{A}] = \frac{\ell}{16 \pi G} \int_\mathrm{AdS} \text{tr} \left( \text{A} \wedge \text{d}\text{A} + \frac{2}{3} \, \text{A} \wedge \text{A} \wedge \text{A} \right) ,
\end{equation}
and an analogous expression for $\bar{\text{A}}$. Here $\ell$ is the AdS radius, while $(\text{A} = A_\mu \text{d}x^\mu,\bar{\text{A}} = \bar{A}_\mu \text{d}x^\mu)$ denote gauge connections valued in the two copies of $\mathfrak{sl}(s,\mathbb{R})$. For concreteness, and to avoid overly lengthy expressions, we focus on the case $s=3$. All results, however, can be extended in a straightforward way to arbitrary value of $s$ along the lines of \cite{Henneaux:2010xg,Gaberdiel:2011wb,Campoleoni:2011hg}.

In the $\mathfrak{sl}(3,\mathbb{R})$ case, we adopt the following basis:
\begin{subequations} \label{eq. sl3basis}
    \begin{align}
    &[L_i,L_j] = \left(i-j\right) L_{i+j} \, ,\\
    &[L_i,W_n] = \left(2i-n\right) W_{i+n} \, ,\\
    &[W_n,W_m] = - \frac{1}{3} \left(n-m\right) \left(2n^2 + 2m^2 - n m - 8\right) L_{n+m} \, ,
    \end{align}
\end{subequations}
where $-1 \leq i,j \leq 1$ and $-2 \leq n,m \leq 2$. We work with bulk coordinates $x^\mu = (z, t, \theta)$, where $z \geq 0$ is the radial coordinate such that the boundary lies at $z=0$, $t$ denotes time, and $\theta \sim \theta + 2\pi$ is the angular coordinate on the asymptotic circle. For convenience, we shall also make frequent use of the boundary light-cone parametrization, $x^\pm = \theta \pm \frac{1}{\ell} t$.

\subsection{Gravity} \label{subsec. gravity}

We begin by reviewing the anomalous relaxations\footnote{The terminology anomalous will be explained by analyzing the associated variational principle in Sect.~\ref{sec. VP}.} discussed in \cite{Alessio:2020ioh,Campoleoni:2022wmf,Ciambelli:2023ott,Delfante:2024npo}, which can be recast in terms of the following boundary conditions for $s=2$:
\begin{subequations} \label{eq. s2solspace1}
    \begin{align}
        \begin{split}
        a_+ &= - \text{e}^{\phi_+} L_1 + e^{-\phi_+} \left( \ell_+ - \left( K_+ \right)^2 - \partial_+ K_+ \right) L_{-1}\\
        &\quad + \left( 2 K_+ + \partial_+ \phi_+ \right) L_0 \, ,\end{split}\\
        a_- &= - e^{-\phi_+} \partial_- K_+ \, L_{-1} + \partial_- \phi_+ \, L_0 \, ,
    \end{align}
\end{subequations}
and
\begin{subequations} \label{eq. s2solspace2}
    \begin{align}
        \begin{split}
        \bar{a}_- &= - \text{e}^{\phi_-} L_{-1} + e^{-\phi_-} \left( \ell_- - \left( K_- \right)^2 - \partial_- K_- \right) L_{1}\\
        &\quad - \left( 2 K_- + \partial_- \phi_- \right) L_0 \, ,\end{split}\\
        \bar{a}_+ &= - e^{-\phi_-} \partial_+ K_- \, L_{1} - \partial_+ \phi_- \, L_0 \, .
     \end{align}
\end{subequations}
On shell, the flatness conditions 
$\text{d}\text{A} + \text{A} \wedge \text{A} \approx 0$ imply that the BH currents $\ell_\pm$ satisfy chirality constraints, $\partial_\mp \ell_\pm \approx 0$, while $K_\pm$ and $\phi_\pm$ remain arbitrary functions on the boundary. The above expressions \eqref{eq. s2solspace1}-\eqref{eq. s2solspace2} rely on the radial gauge choice
\begin{subequations} \label{eq. RadialGaugeFixing}
    \begin{align}
    &A_z = b^{-1}(z) \partial_z b(z) \, , &&\bar{A}_z = \bar{b}^{-1}(z) \partial_z \bar{b}(z) \, ,\label{eq. RadialGaugeFixing1}\\
    &A_\pm = b^{-1}(z) \, a_\pm \, b(z) \, , &&\bar{A}_\pm = \bar{b}^{-1}(z) \, \bar{a}_\pm \, \bar{b}(z) \, ,\label{eq. RadialGaugeFixing2}
    \end{align}
\end{subequations}
with $b(z), \bar{b}(z) \in SL(2,\mathbb{R})$. The fixation~\eqref{eq. RadialGaugeFixing1} can always be imposed off shell in a neighborhood of the boundary~\cite{Campoleoni:2010zq}.

The splitting of the solution space~\eqref{eq. RadialGaugeFixing2} shows that the radial coordinate can be completely gauged away in CS formulation. In particular, the radial dependence cancels identically in the on-shell evaluation of the action~\eqref{eq. actionCS}, thanks to the cyclicity of the trace. As we shall see in the next Section, the same holds for the asymptotic charges, since they are constructed directly from the action. In contrast, the radial coordinate retains a genuine role in the second-order formalism, where it acquires a clear geometric interpretation. We shall return to this point in~\ref{sec. RC}, analyzing two distinct choices of radial group elements. Actually, depending on how these transformations shape the nature of the holographic radial coordinate, namely spacelike in the Fefferman–Graham (FG) gauge~\cite{FG1} or null in the Bondi gauge~\cite{Bondi:1960jsa,Sachs:1961zz}, our boundary conditions~\eqref{eq. s2solspace1}–\eqref{eq. s2solspace2} admit two respective significant reductions.

More concretely, in the spacelike case, if we restrict the field content as $\{\phi_\pm = \varphi,K_\pm = k_\pm - \partial_\pm \varphi\}$, we recover the CS connections of the Weyl--Fefferman--Graham (WFG) formulation~\cite{Ciambelli:2019bzz,Ciambelli:2023ott}. In this setting, $\varphi(x^+,x^-)$ plays the role of the conformal factor of the boundary metric, while $k_\pm(x^+,x^-)$ correspond to the boundary Weyl connection. Alternatively, if we reduce the arbitrary functions as $\{\phi_\pm = \varphi \pm \zeta,K_\pm = \mp \partial_\pm \zeta\}$ in the null case, the resulting connections reproduce those of the covariant Bondi (CB) formulation~\cite{Campoleoni:2022wmf} for asymptotically AdS spacetimes. Here, $\varphi(x^+,x^-)$ again serves as the boundary conformal factor, while $\zeta(x^+,x^-)$ is the connection associated to the Lorentz boost of the boundary zweibein.

In \ref{sec. RC}, we illustrate in detail how our solution space~\eqref{eq. s2solspace1}-\eqref{eq. s2solspace2} induces gauge relaxations of the WFG and CB types. A more systematic and quantitative holographic interpretation of this general framework will be left for future work.

\subsection{Higher Spin Gravity}

To proceed, we now identify the $SL(2,\mathbb{R})$ gauge transformation mapping the new boundary conditions~\eqref{eq. s2solspace1}--\eqref{eq. s2solspace2} to the Brown--Henneaux (BH) ones, where only the (anti-)chiral functions~$\ell_\pm$ are switched on. It is given by
\begin{subequations} \label{eq. s2largegaugetransfo}
    \begin{align}
    \text{a} &= B^{-1} \left[ \left( - L_1 + \ell_+ L_{-1} \right) \text{d}x^+ \right] B + B^{-1} \text{d} B \, ,\\
    \bar{\text{a}} &= \bar{B}^{-1} \left[ \left( - L_{-1} + \ell_- L_{1} \right) \text{d}x^- \right] \bar{B} + \bar{B}^{-1} \text{d} \bar{B} \, ,
    \end{align}
\end{subequations}
with group elements
\begin{subequations} \label{eq. s2gengroupelem}
    \begin{align}
    B(x^+,x^-) &= \text{exp} \left( - K_+ L_{-1} \right) \, \text{exp} \left(  \phi_+ L_0 \right) ,\\
    \bar{B}(x^+,x^-) &= \text{exp} \left( - K_- L_{1} \right) \, \text{exp} \left( - \phi_- L_0 \right) .
    \end{align}
\end{subequations}
This transformation thus generates additional gravitational content beyond the BH sector through the arbitrary functions $K_\pm$ and $\phi_\pm$. As we shall see in the next Section, it corresponds to a large gauge transformation.

In the spin-$3$ case, the natural analogue of the Brown--Henneaux boundary conditions is provided by the highest-weight ones \cite{Coussaert:1995zp}, which implement the Drinfeld--Sokolov (DS) reduction~\cite{Drinfeld:1984qv,BALOG199076},
\begin{subequations} \label{eq. DrinfeldSokolov}
    \begin{align}
    \text{a} &= \left( - L_1 + \ell_+ L_{-1} + \frac{1}{4} w_+ W_{-2} \right) \text{d}x^+ \, ,\\
    \bar{\text{a}} &= \left( - L_{-1} + \ell_- L_{1} + \frac{1}{4} w_- W_{2} \right) \text{d}x^- \, .
    \end{align}
\end{subequations}
It introduces an additional pair of functions, $w_\pm$, subject to the on-shell chirality constraints $\partial_\mp w_\pm \approx 0$. Since the two copies have the same structure up to minor redefinitions, we will for conciseness focus on the first copy, indicating the necessary modifications for the second along the way.

The higher-spin counterpart of the large gauge transformation~\eqref{eq. s2largegaugetransfo} is obtained by extending the group element~\eqref{eq. s2gengroupelem} to its $SL(3,\mathbb{R})$ generalization:
\begin{equation} \label{eq. BMphi}
    B = M(K_+^L,K_+^W) \exp \left( \phi_+^L L_0 + \phi_+^W W_0 \right) .
\end{equation}
Here, the contributions along the zero modes (spanning the Cartan subalgebra) admit a straightforward generalization and are labeled by superscripts $L$ and $W$, distinguishing whether the corresponding boundary field is of gravitational or higher-spin origin, respectively. By contrast, the factor $M = M(K_+^L,K_+^W)$, which collects the contributions aligned with the lowest-weight modes of the gauge algebra, requires a more delicate treatment.

Actually, a prescription was put forward in~\cite{Li:2015osa}, relying on the Miura transformation~\cite{Fateev:1987zh}:
\begin{equation} \label{eq. Miura}
    M \left( - L_1 + \vec{\mathcal{K}} \right) M^{-1} + M \text{d} M^{-1} = - L_1 + \vec{\mathcal{Q}} \, ,
\end{equation}
with $\vec{\mathcal{K}} = \vec{\mathcal{K}}(K^L_+,K^W_+)$ and $\vec{\mathcal{Q}} = \vec{\mathcal{Q}}(K^L_+,K^W_+)$. Structurally, the 
Miura map~\eqref{eq. Miura} associates to a Cartan element 
\(\vec{\mathcal{K}} \in \mathfrak{sl}(3,\mathbb{R})\) a unique group element \(M\) generated solely by negative modes. The presence of \(\vec{\mathcal{K}}\) is motivated by analogy with the spin-2 case (cf.~Eq.~\eqref{eq. s2solspace1}), while the resulting image \(\vec{\mathcal{Q}}\) lies entirely along lowest-weight directions, again mirroring the gravitational construction. In fact, the analysis of~\cite{Li:2015osa} was conducted in the restricted FG framework, where only the conformal factor $\varphi^L$ is switched on ($k_\pm = 0$). This led to a higher-spin generalization of holographic structures already known from pure gravity, such as the Weyl anomaly~\cite{Henningson:1998gx}. Within that setting, a conformal factor $\varphi^W$ was also introduced for the spin-3 field, playing the role of the radial rescaling symmetry familiar from the spin-2 gauge. This is the rationale behind our choice in~\eqref{eq. BMphi}. For further details we refer to their original work. Building on this idea, we continue their analysis by determining the associated asymptotic symmetries and even extend their prescription \eqref{eq. Miura} to include the anomalous contributions $(K_\pm^{L,W})$.

Following these steps, we deduce that
\begin{equation}
    M = \exp \left[ - K_+^L L_{-1} - \frac{2}{3} K_+^W W_{-1} + \frac{1}{6} \left( \partial_+ K_+^W + 4 K_+^L K_+^W \right) W_{-2} \right] ,
\end{equation}
so that the generalization of the new boundary conditions~\eqref{eq. s2solspace1} to the spin-$3$ coupling is given by
\begin{align*}
    &a_+ = - \text{e}^{\phi_+^L} \cosh(2\phi_+^W) L_1 - \frac{\text{e}^{-\phi_+^L}}{3} \Big[ \cosh(2\phi_+^W) \Big(-3\ell_+ + 3 (K_+^L)^2\\
    &+ 4 (K_+^W)^2 + 3 \partial_+ K_+^L\Big) - 4 \sinh(2\phi_+^W) \Big( \partial_+ K_+^W + 4 K_+^L K_+^W \Big) \Big] L_{-1}\\
    &+ \Big(2 K_+^L + \partial_+ \phi_+^L\Big) L_0 - \text{e}^{\phi_+^L} \sinh(2\phi_+^W) W_1 + \Big(2 K_+^W + \partial_+ \phi_+^W\Big) W_0\\
    &+ \frac{\text{e}^{-\phi_+^L}}{3} \Big[ \sinh(2\phi_+^W) \Big(-3\ell_+ + 3 (K_+^L)^2 + 4 (K_+^W)^2 + 3 \partial_+ K_+^L\Big)\\
    &- 4 \cosh(2\phi_+^W) \Big(\partial_+ K_+^W + 4 K_+^L K_+^W\Big) \Big] W_{-1} + \frac{\text{e}^{-2\phi_+^L}}{54} \Big[ \frac{27}{2} w_+\\
    &+ 2 K_+^W \Big( 18 \ell_+ + 90 (K_+^L)^2 - 8 (K_+^W)^2 + 9 \partial_+ K_+^L\Big) + 90 K_+^L \partial_+ K_+^W\\
    &+ 9 \partial_+^2 K_+^W \Big] W_{-2} \, , \stepcounter{equation}\tag{\theequation}\label{eq. s3solspace1}
\end{align*}
and
\begin{align*}
    &a_- = \frac{\text{e}^{-\phi_+^L}}{3} \Big[ - 3 \cosh(2\phi_+^W) \partial_- K_+^L + 2 \sinh(2\phi_+^W) \partial_- K_+^W \Big] L_{-1}\\
    &+ \partial_- \phi_+^L L_0 + \partial_- \phi_+^W W_0 + \frac{\text{e}^{-\phi_+^L}}{3} \Big[ 3 \sinh(2\phi_+^W) \partial_- K_+^L\\
    &- 2 \cosh(2\phi_+^W) \partial_- K_+^W \Big] W_{-1} + \frac{\text{e}^{-2\phi_+^L}}{6} \Big[ \partial_+ \partial_- K_+^W\\
    &+ 2 K_+^W \partial_- K_+^L + 6 K_+^L \partial_- K_+^W \Big] W_{-2} \, .\stepcounter{equation}\tag{\theequation}\label{eq. s3solspace2}
\end{align*}
The second copy is obtained analogously under the replacements $\bar{a}_\pm = a_\mp$, $L_{i\neq 0} \leftrightarrow L_{-i}$, $L_0 \to - L_0$, $W_{n\neq 0} \leftrightarrow W_{-n}$, $W_0 \to - W_0$, $\partial_\pm \leftrightarrow \partial_\mp$, and $\psi_+ \leftrightarrow \psi_-$, where we have denoted the field content by $\psi_\pm = \{\ell_\pm,w_\pm,K^{L,W}_\pm,\phi^{L,W}_\pm\}$.

%%%%%%%%%%%%%%%%%%%%%%%%%%%%%%%%%%%%%%

\section{Asymptotic Symmetries} \label{sec. AS}

\subsection{Residual Symmetries}

As repeatedly emphasized throughout this Letter, the CS action \eqref{eq. actionCS} is invariant under gauge transformations of the form
\begin{equation} \label{eq. SLnGaugeTransfoA}
    \text{A} \to U^{-1} \text{A} \, U + U^{-1} \text{d} U \, ,
\end{equation}
with $U = \text{exp}(\lambda)$ and $\lambda \in \mathfrak{sl}(3,\mathbb{R})$. At the infinitesimal level, it reads
\begin{equation} \label{eq. gauge sym CS}
    \delta_\lambda \text{A} = \text{d}\lambda + [\text{A},\lambda] \, .
\end{equation}
The residual symmetries compatible with the boundary conditions \eqref{eq. s3solspace1}–\eqref{eq. s3solspace2} are generated by $\lambda = \sum_i \epsilon^i L_i + \sum_n \sigma^n W_n$, with explicit components
\begin{align*}
    &\epsilon^1 = \frac{\text{e}^{\phi_+^L}}{3} \cosh(2\phi_+^W) \Big[ 3 Y^+ + 8 K_+^W \chi^+ + 3 \tanh(2\phi_+^W) \Big(\partial_+ \chi^+\\
    & - 4 K_+^L \chi^+\Big) \Big] \, ,\\
    %%%%%%
    &\epsilon^0 = \omega_+^L + \frac{1}{6} \Big[ 3 (\partial_+ Y^+ - \partial_- Y^-) - 2 K_+^L (3 Y^+ + 16 K_+^W \chi^+)\\
    & + 2 K_-^L (3 Y^- + 16 K_-^W \chi^-) - 4 (K_+^W \partial_+ \chi^+ + K_-^W \partial_- \chi^-)\\
    & - 8 (\partial_+ K_+^W \chi^+ - \partial_- K_-^W \chi^-) \Big] \, ,\\
    %%%%%%
    &\epsilon^{-1} = \frac{\text{e}^{-\phi_+^L}}{54} \Big[ 3 \sinh(2\phi_+^W) \Big( 12 H_+^W + 72 (K_+^L)^3 \chi^+ + 12 K_+^W \partial_+ Y^+\\
    & - ( 54 (K_+^L)^2 + 8 (K_+^W)^2 - 30 \ell_+ ) \partial_+ \chi^+ - 2 K_+^L ( 48 K_+^W Y^+\\
    & + 36 \ell_+ \chi^+ + 80 (K_+^W)^2 \chi^+ - 9 \partial_+^2 \chi^+ ) - 3 \partial_+^3 \chi^+ + 12 \chi^+ \partial_+ \ell_+\\
    & - 4 ( 3 Y^+ + 8 K_+^W \chi^+ ) \partial_+ K_+^W \Big) + \cosh(2\phi_+^W) \Big( 72 (K_+^W)^2 Y^+\\
    & + 64 (K_+^W)^3 \chi^+ + 54 (K_+^L)^2 (Y^+ + 8 K_+^W \chi^+) + 36 K_+^W (4 \ell_+ \chi^+\\
    & - \partial_+^2 \chi^+) - 18 K_+^L (3 \partial_+ Y^+ - 8 \chi^+ \partial_+ K_+^W) - 9 (6 H_+^L + 6 \ell_+ Y^+\\
    & - 12 w_+ \chi^+ - 3 \partial_+^2 Y^+ + 4 \partial_+ \chi^+ \partial_+ K_+^W ) \Big) \Big] \, ,\\
    %%%%%%
    &\sigma^2 = \text{e}^{2\phi_+^L} \chi^+ \, ,\\
    %%%%%%
    &\sigma^1 = \frac{\text{e}^{\phi_+^L}}{3} \Big[ \sinh(2\phi_+^W) \Big(3 Y^+ + 8 K_+^W \chi^+\Big) + 3 \cosh(2\phi_+^W) \Big(\partial_+ \chi^+\\
    & - 4 K_+^L \chi^+\Big) \Big] \, ,\\
    %%%%%%
    &\sigma^0 = \omega_W + K_-^W Y^- - K_+^W Y^+ - \frac{4}{3} \Big[ (K_-^W)^2 \chi^- + (K_+^W)^2 \chi^+ \Big]\\
    & + \frac{1}{4} \Big[ 12 \Big( (K_-^L)^2 \chi^- + (K_+^L)^2 \chi^+ \Big) - 4 \Big( \ell_- \chi^- + \ell_+ \chi^+ \Big)\\
    &- 6 \Big( K_-^L \partial_- \chi^- + K_+^L \partial_+ \chi^+ \Big) + \partial_-^2 \chi^- + \partial_+^2 \chi^+ \Big] \, . \stepcounter{equation}\tag{\theequation}\label{eq. s3residualsymm}
\end{align*}
The field parameters are the (anti-)chiral functions $\{Y^\pm(x^\pm),\chi^\pm(x^\pm)\}$, together with the arbitrary fields $\{\omega_\pm^{L,W}(x^+,x^-),H_\pm^{L,W}(x^+,x^-)\}$. The last two gauge components are not independent but satisfy
\begin{subequations}
\begin{align}
    &\sigma^{-1} = \frac{1}{a_+^{\ell_1}} \Big[ \frac{1}{3} (\delta_\lambda a_+^{w_0} - \partial_+ \sigma^0) + a_+^{\ell_{-1}} \sigma^1 - a_+^{w_1} \epsilon^{-1} + a_+^{w_{-1}} \epsilon^1 \Big] \, ,\\
    %%%%%%
    &\sigma^{-2} = \frac{1}{4 a_+^{\ell_{1}}} \Big[ \delta_\lambda a_+^{w_{-1}} - \partial_+ \sigma^{-1} + 2 a_+^{\ell_{-1}} \sigma^0 - a_+^{\ell_0} \sigma^{-1} + a_+^{w_{-1}} \epsilon^0 \notag\\
    & - a_+^{w_0} \epsilon^{-1} + 4 a_+^{w_{-2}} \epsilon^1 \Big] \, , \stepcounter{equation}\tag{\theequation}
\end{align}
\end{subequations}
where we decomposed $a_+ = \sum_i a_+^{\ell_i} L_i + \sum_n a_+^{w_n} W_n$. The barred copy of the gauge parameters, $\bar{\lambda}$, is obtained by the same substitutions as for the CS connections (cf.~Eq.~\eqref{eq. s3solspace2}), with the additional replacement $\xi_+ \leftrightarrow \xi_-$, where we denote the set of independent parameters as $\xi_\pm = \{Y^\pm,\chi^\pm,H_\pm^{L,W},\omega_\pm^{L,W}\}$.

Under these gauge laws, the field contents $\psi_\pm$ vary as
\begin{subequations}
    \begin{align}
        \begin{split}
        \delta_\Lambda \ell_\pm &= Y^\pm \partial_\pm \ell_\pm + 2 \ell_\pm \partial_\pm Y^\pm - 2 \chi^\pm \partial_\pm w_\pm - 3 w_\pm \partial_\pm \chi^\pm\\
        &\quad + \frac{1}{2} \partial_\pm^3 Y^\pm \, ,
        \end{split}\\
        \begin{split}
        \delta_\Lambda w_\pm &= Y^\pm \partial_\pm w_\pm + 3 w_\pm \partial_\pm Y^\pm + \frac{1}{12} \Big( (2 \chi^\pm \partial_\pm^3\\
        &\quad + 9 \partial_\pm \chi^\pm \partial_\pm^2 + 15 \partial_\pm^2 \chi^\pm \partial_\pm + 10 \partial_\pm^3 \chi^\pm) \ell_\pm\\
        &\quad + 32 \ell_\pm (\chi^\pm \partial_\pm \ell_\pm + \ell_\pm \partial_\pm \chi^\pm) + \frac{1}{2} \partial_\pm^5 \chi^\pm \Big) \, ,
        \end{split}\\
        \delta_\Lambda \phi^{L,W}_\pm &= \sigma^{L,W}_\pm \, ,\\
        \delta_\Lambda K^{L,W}_\pm &= H^{L,W}_\pm \, ,
    \end{align}
\end{subequations}
where we have labeled $\Lambda = (\lambda,\bar{\lambda})$ and redefined
\begin{subequations} \label{eq. sigmaomega}
    \begin{align}
        \begin{split}
        \sigma_\pm^L &= \omega_\pm^L + \frac{1}{6} \Big( K_-^L (6 Y^- + 32 K_-^W \chi^-) + K_+^L (6 Y^+\\
        &\quad + 32 K_+^W \chi^+) - 3 (\partial_+ Y^+ + \partial_- Y^-) - 4 (K_-^W \partial_- \chi^-\\
        &\quad - K_+^W \partial_+ \chi^+) + 8 (\chi^- \partial_- K_-^W + \chi^+ \partial_+ K_+^W) \Big) \, ,
        \end{split}\\
        \begin{split}
        \sigma_\pm^W &= \omega_\pm^W + K_-^W Y^- + K_+^W Y^+ - \frac{4}{3} \Big( (K_-^W)^2 \chi^- - (K_+^W)^2 \chi^+ \Big)\\
        &\quad + \frac{1}{4} \Big( 6 K_-^L (2 K_-^L - \partial_-) \chi^- - 6 K_+^L (2 K_+^L - \partial_+) \chi^+\\
        &\quad + \partial_-^2 \chi^- - \partial_+^2 \chi^+ \Big) + \ell_- \chi^+ - \ell_- \chi^- \, ,
        \end{split}
    \end{align}
\end{subequations}
so that the usual relation between Weyl rescalings and $\mathfrak{sl}(3,\mathbb{R})$ gauge transformations along Cartan directions \cite{Rooman:2000zi,Banados:2004nr} takes the convenient form
\begin{equation}
    \epsilon^0 - \bar{\epsilon}^0 = \omega_+^L + \omega_-^L \, , \qquad \sigma^0 - \bar{\sigma}^0 = \omega_+^W + \omega_-^W \, .
\end{equation}
In fact, the appearance of $K_\pm^{L,W}$ in the redefinitions \eqref{eq. sigmaomega} suggests a geometric covariantization with respect to these fields, compared to the standard DS boundary conditions~\eqref{eq. DrinfeldSokolov}. This feature had already been anticipated in the WFG and CB restrictions below Eq.~\eqref{eq. RadialGaugeFixing2}. We will return to some additional geometric remarks in \ref{sec. RC}.

\subsection{Asymptotic Charges}

The presymplectic potential associated with the CS action \eqref{eq. actionCS} is defined through the following on-shell variation:
\begin{subequations} \label{eq. ThetaCS}
    \begin{align}
    &\delta S_\mathrm{CS}[\text{A}] \approx \int_\mathrm{AdS} \text{d}
    \Theta[\text{A},\delta \text{A}] \, ,\\
    &\Theta[\text{A},\delta \text{A}] = \frac{\ell}{16 \pi G} \text{tr} \left( \text{A} \wedge \delta \text{A} \right) .
    \end{align}
\end{subequations}
In the Iyer--Wald formalism~\cite{Lee:1990nz,Wald:1993nt,Iyer:1994ys}, the surface charge densities are obtained by evaluating on-shell the presymplectic form 
$\omega[\delta \text{A},\delta \text{A}] = \delta \Theta[\text{A},\delta \text{A}]$ along the gauge symmetries~\eqref{eq. gauge sym CS},
\begin{equation} \label{eq. 2ndNoether}
    \omega_\lambda:= \omega[\delta_\lambda \text{A},\delta \text{A}] = \delta_\lambda \Theta[\text{A},\delta \text{A}] - \delta \Theta[\text{A},\delta_\lambda \text{A}] \approx \text{d}\text{k}_\lambda \, .
\end{equation}
The charge of the full theory~\eqref{eq. CS_generic_split} then follows as
\begin{equation} \label{eq. generic full charge}
    \delta \text{Q}_\Lambda = \delta \text{q}_\lambda - \delta \bar{\text{q}}_{\bar{\lambda}} \, ,
\end{equation}
with
\begin{equation} \label{eq. generic CS charge}
    \delta \text{q}_\lambda \approx \int_{\mathcal{S}} \text{k}_\lambda \, , \qquad \text{k}_\lambda = - \frac{\ell}{8 \pi G} \text{tr} \left( \lambda \, \delta \text{A} \right) ,
\end{equation}
where $\mathcal{S}$ denotes the boundary circle at infinity. This result agrees with the charge first derived in \cite{Banados:1994tn}. An analogous expression holds for the second copy.

Specializing to the solution space~\eqref{eq. s3solspace1}--\eqref{eq. s3solspace2} and the residual symmetries~\eqref{eq. s3residualsymm}, the asymptotic surface charge~\eqref{eq. generic full charge} takes the explicit form
\begin{equation} \label{eq. s3asymptoticcharge}
    \begin{split}
    &\text{Q}_\Lambda = \frac{\ell}{8 \pi G} \int \text{d}\theta \, \Big[ Y^+ \ell_+ - Y^- \ell_- + \chi^+ w_+ - \chi^- w_-\\
    &\quad + \big( H_+^L \phi_+^L - K_+^L \sigma_+^L \big) - \big( H_-^L \phi_-^L - K_-^L \sigma_-^L \big)\\
    &\quad + \frac{4}{3} \big( H_+^W \phi_+^W - K_+^W \sigma_+^W \big) - \frac{4}{3} \big( H_-^W \phi_-^W - K_-^W \sigma_-^W \big)\\
    &\quad - \frac{1}{2} \big( \sigma_+^L \partial_\theta \phi_+^L - \sigma_-^L \partial_\theta \phi_-^L \big) - \frac{2}{3} \big( \sigma_+^W \partial_\theta \phi_+^W - \sigma_-^W \partial_\theta \phi_-^W \big) \Big] \, .
    \end{split}
\end{equation}
Notice that the latter has been trivially integrated.\footnote{In fact, in three dimensions, integrable slicings for the gravitational and higher-spin charges can always be constructed~\cite{Ruzziconi:2020wrb,Geiller:2021vpg}, reflecting the absence of propagating local degrees of freedom for $s > 1$.} Furthermore, we emphasize that these charges~\eqref{eq. s3asymptoticcharge} are radially independent and therefore finite at asymptotic infinity, without requiring any holographic renormalization procedure.

\subsection{Charge Algebra}

We define the Poisson bracket as $\{\text{Q}_{\Lambda_1},\text{Q}_{\Lambda_2}\} = \delta_{\Lambda_2} \text{Q}_{\Lambda_1}$. Then, the first line of \eqref{eq. s3asymptoticcharge} reproduces the conserved Brown--Henneaux and Drinfeld--Sokolov charges, which generate two copies of the centrally extended classical Zamolodchikov $\mathcal{W}_3$ algebra~\cite{Zamolodchikov:1985wn,Henneaux:2010xg,Campoleoni:2010zq}:
\begin{subequations} \label{eq. W3algebra}
    \begin{align}
        \text{i} \{\text{Q}_{\Lambda^{Y^\pm}_{p}}, \text{Q}_{\Lambda^{Y^\pm}_{q}}\} &= (p-q) \text{Q}_{\Lambda^{Y^\pm}_{p+q}} + \frac{c}{12} p^3 \delta_{p+q,0} \, ,\\
        \text{i} \{\text{Q}_{\Lambda^{Y^\pm}_{p}}, \text{Q}_{\Lambda^{\chi^\pm}_{q}}\} &= (2p-q) \text{Q}_{\Lambda^{\chi^\pm}_{p+q}} \, ,\\
        \begin{split}
        \text{i} \{\text{Q}_{\Lambda^{\chi^\pm}_{p}}, \text{Q}_{\Lambda^{\chi^\pm}_{q}}\} &= (p-q) (2p^2 + 2q^2 - pq) \text{Q}_{\Lambda^{Y^\pm}_{p+q}}\\
        &\quad + \frac{96}{c} (p-q) \text{H}_{\Lambda^{Y^\pm}_{p+q}} + \frac{c}{12} p^5 \delta_{p+q,0} \, ,
        \end{split}
    \end{align}
\end{subequations}
where $c = \tfrac{3\ell}{2G}$ is the BH central charge \cite{Brown:1986nw}, and we introduced the quadratic term $\text{H}_{\Lambda^{Y^\pm}_{p}} = \sum_{q \in \mathbb{Z}} \text{Q}_{\Lambda^{Y^\pm}_{p+q}} \text{Q}_{\Lambda^{Y^\pm}_{-q}}$. In the above commutators, the (anti-)chiral parameters are expanded in Fourier modes as $Y^\pm \sim \text{e}^{\text{i}p x^\pm}$ and $\chi^\pm \sim \text{e}^{\text{i}p x^\pm}$. The notation $\Lambda_p^{Y^\pm}$ indicates gauge parameters where only the Fourier modes of $Y^\pm$ are switched on, with all others set to zero.

The remaining contributions in~\eqref{eq. s3asymptoticcharge} constitute the main novelty of our analysis. These charges depend on the arbitrary boundary fields $\{K_\pm^{L,W},\phi_\pm^{L,W}\}$ and are non-conserved, hence our previous designation of them as anomalous charges.\footnote{Again, the anomalies that these charges induce will become manifest when examining the on-shell variation of the action \eqref{eq. CS_generic_split} in Sect.~\ref{sec. VP}.} These obey the following non-trivial commutation relations:
\begin{subequations} \label{eq. anomalalgebra}
    \begin{align}
        \text{i} \{\text{Q}_{\Lambda^{\sigma_\pm^L}_{p,q}}, \text{Q}_{\Lambda^{H_\pm^L}_{r,s}}\} &= \pm \frac{c}{12} \text{e}^{2 \frac{\text{i}}{\ell} (q+s) t} \delta_{p+r,q+s} \, ,\\
        \text{i} \{\text{Q}_{\Lambda^{\sigma_\pm^L}_{p,q}}, \text{Q}_{\Lambda^{\sigma_\pm^L}_{r,s}}\} &= \mp \frac{c}{24} (r+s) \text{e}^{2 \frac{\text{i}}{\ell} (q+s) t} \delta_{p+r,q+s} \, ,\\
        \text{i} \{\text{Q}_{\Lambda^{\sigma_\pm^W}_{p,q}}, \text{Q}_{\Lambda^{H_\pm^W}_{r,s}}\} &= \pm \frac{c}{9} \text{e}^{2 \frac{\text{i}}{\ell} (q+s) t} \delta_{p+r,q+s} \, ,\\
        \text{i} \{\text{Q}_{\Lambda^{\sigma_\pm^W}_{p,q}}, \text{Q}_{\Lambda^{\sigma_\pm^W}_{r,s}}\} &= \mp \frac{c}{18} (r+s) \text{e}^{2 \frac{\text{i}}{\ell} (q+s) t} \delta_{p+r,q+s} \, ,
    \end{align}
\end{subequations}
where the arbitrary boundary parameters are expanded as $\sigma_\pm^{L,W} \sim \text{e}^{\text{i}p x^+} \text{e}^{\text{i}q x^-}$ and $H_\pm^{L,W} \sim \text{e}^{\text{i}p x^+} \text{e}^{\text{i}q x^-}$. Notably, the explicit time dependence in the central terms precisely reproduces the structure identified in \cite{Alessio:2020ioh,Campoleoni:2022wmf,Ciambelli:2023ott} in the gravity sector. Furthermore, these centrally extended Abelian sectors form a one-parameter family of algebras, with the parameter given by the time \(t\) at which the surface charges are evaluated.

%%%%%%%%%%%%%%%%%%%%%%%%%%%%%%%%%%%%%%

\section{Variational Principle} \label{sec. VP}

\subsection{Boundary Term}

Adding the Coussaert--Henneaux--van~Driel boundary term~\cite{Coussaert:1995zp} to the action \eqref{eq. CS_generic_split} with DS boundary conditions \eqref{eq. DrinfeldSokolov} ensures a well-defined variational principle:
\begin{equation} \label{eq. CHvD}
    S_\mathrm{tot} = S + S_\mathrm{CHvD} \, , \quad S_\mathrm{CHvD} = - \frac{1}{16 \pi G} \int \text{d}^2x \, \text{tr} \left( A_\theta^2 + \bar{A}_\theta^2 \right) ,
\end{equation}
so that $\delta S_\mathrm{tot} \approx 0$. This prescription, however, ceases to be sufficient once the more general connections~\eqref{eq. s3solspace1}-\eqref{eq. s3solspace2} are considered. In that case, the action must be supplemented by the additional boundary contribution
\begin{align*}
    &S_{\mathrm{bdy}} = \frac{1}{16 \pi G} \int \text{d}^2x \, \Big[ (\partial_\theta + \ell \partial_t) \phi_-^L (K_\theta^L - \ell K_t^L)\\
    & + (\partial_\theta - \ell \partial_t) \phi_+^L (K_\theta^L + \ell K_t^L) + \frac{\ell}{2} ( \partial_t \phi_-^L \partial_\theta \phi_-^L - \partial_t \phi_+^L \partial_\theta \phi_+^L)\\
    & + \frac{1}{2} \big( (\partial_\theta \phi_-^L)^2 + (\partial_\theta \phi_+^L)^2 \big) + \frac{4}{3} (\partial_\theta + \ell \partial_t) \phi_-^W (K_\theta^W - \ell K_t^W)\\
    & + \frac{4}{3} (\partial_\theta - \ell \partial_t) \phi_+^W (K_\theta^W + \ell K_t^W) + \frac{2\ell}{3} ( \partial_t \phi_-^W \partial_\theta \phi_-^W - \partial_t \phi_+^W \partial_\theta \phi_+^W)\\
    & + \frac{2}{3} \big( (\partial_\theta \phi_-^W)^2 + (\partial_\theta \phi_+^W)^2 \big) \Big] \, , \stepcounter{equation}\tag{\theequation} \label{eq. s3 bdy term}
\end{align*}
where we have parametrize $K_\pm = \tfrac{1}{2} (K_\theta \pm \ell K_t)$, in analogy with the WFG and CB interpretations. Nevertheless, even with this extension \eqref{eq. s3 bdy term}, the on-shell variation of the total action remains non-integrable, and takes the form
\begin{align*}
    &\Theta^r_\mathrm{tot} = \frac{\ell}{16 \pi G} \Big[ \frac{1}{\ell} \partial_\theta \Big( K_\theta^L \delta (\phi_+^L + \phi_-^L) + \ell K_t^L \delta (\phi_+^L - \phi_-^L)\\
    & - \frac{\ell}{2} (\phi_+^L \partial_t \delta \phi_+^L - \phi_-^L \partial_t \delta \phi_-^L) - \frac{2\ell}{3} (\phi_+^W \partial_t \delta \phi_+^W - \phi_-^W \partial_t \delta \phi_-^W)\\
    & + \frac{4}{3} K_\theta^W \delta (\phi_+^W + \phi_-^W) + \frac{4\ell}{3} K_t^W \delta (\phi_+^W - \phi_-^W) \Big)\\
    & - \partial_t \Big( K_\theta^L \delta (\phi_+^L - \phi_-^L) + \ell K_t^L \delta (\phi_+^L + \phi_-^L) - \frac{1}{2} ( \phi_+^L \partial_\theta \delta \phi_+^L\\
    & - \phi_-^L \partial_\theta \delta \phi_-^L ) + \frac{4}{3} K_\theta^W \delta (\phi_+^W - \phi_-^W) + \frac{4\ell}{3} K_t^W \delta (\phi_+^W + \phi_-^W)\\
    & - \frac{2}{3} ( \phi_+^W \partial_\theta \delta \phi_+^W - \phi_-^W \partial_\theta \delta \phi_-^W ) \Big) \Big] \, . \stepcounter{equation}\tag{\theequation} \label{eq. Thetartot}
\end{align*}
It demonstrates that, unless additional constraints are imposed on the solution space, no further boundary term can render the variational problem well-posed~\cite{Alessio:2020ioh,Campoleoni:2022wmf,Ciambelli:2023ott}. It thus reveals the presence of an anomaly in the dual theory along the symmetries generated by $\sigma_\pm^{L,W}$ and $H_\pm^{L,W}$.

\subsection{Symplectic Structure} \label{subsec. ST}

In this Subsection, we stress that the two residual contributions to the on-shell variation of the total action \eqref{eq. Thetartot} are not dynamical, but rather ambiguities intrinsic to the definition of the surface charge~\eqref{eq. generic full charge}-\eqref{eq. generic CS charge}. To make this point clear, it is convenient to phrase the discussion in symplectic language.

The presymplectic structure~\eqref{eq. ThetaCS} admits two classes of redefinitions that leave the equations of motion unchanged:
\begin{equation} \label{eq. IWambi}
    \Theta[\text{A},\delta \text{A}] \to \Theta[\text{A},\delta \text{A}] + \delta \text{B}[\text{A}] + \text{d}\text{C}[\text{A},\delta \text{A}] \, ,
\end{equation}
with $\text{C}$ being skew-symmetric. The first ambiguity amounts to adding a boundary term to the Lagrangian. It does not modify the presymplectic form $\omega$ nor the associated Noether charges~\eqref{eq. 2ndNoether}. The second ambiguity arises because~$\Theta$ enters $\delta S$ as a boundary contribution. This ``corner term'' modifies $\omega_\lambda$ according to
\begin{equation} \label{eq. cornerambi}
    \omega_\lambda \to \omega_\lambda + \text{d} \, \Big( \delta_\lambda \text{C}[\text{A},\delta \text{A}] - \delta \text{C}[\text{A},\delta_\lambda \text{A}] \Big) \, ,
\end{equation}
and thereby shifts the value of the surface charge density~\eqref{eq. 2ndNoether}.

The total presymplectic potential~\eqref{eq. Thetartot} is precisely of this second type, as it involves only total derivatives. It immediately raises an interpretational issue for the associated charges $\{\phi_\pm^{L,W}, K_\pm^{L,W}\}$: they can be absorbed into a redefinition of the presymplectic potential, without any dynamical consequence. This stands in sharp contrast with the DS charges $\{\ell_\pm, w_\pm\}$, which are tied to a genuine boundary term and therefore cannot be kinematically subtracted from the charge value. This subtlety has already been identified as a peculiarity in various contexts, most notably in the WFG and CB gauges of pure gravity \cite{Geiller:2021vpg,Campoleoni:2023fug,Ciambelli:2023ott}. In particular, \cite{Ciambelli:2024vhy} distinguished between two classes of charges: ``dynamical'' charges, associated with constrained degrees of freedom, and ``kinematical'' charges, associated with unconstrained ones. They further showed that these classes display distinct covariance properties under boundary diffeomorphisms in pure gravity. Here we extend this discussion to higher-spin couplings by providing an additional viewpoint to address it.

In fact, the possibility of rewriting such charges as corner terms is a structural feature of the symplectic formalism itself. Already in the sense of the first Noether theorem, the physical status of these charges is dubious: the fields $\{\phi_\pm^{L,W}, K_\pm^{L,W}\}$ and their associated symmetries $\{\sigma_\pm^{L,W}, H_\pm^{L,W}\}$ are unconstrained by the equations of motion and remain completely free on the boundary. This kinematic freedom precisely manifests as a corner ambiguity in the Iyer--Wald formalism~\cite{Lee:1990nz,Wald:1993nt,Iyer:1994ys}. Indeed, since the asymptotic solution space does not restrict these fields, the relation~\eqref{eq. 2ndNoether} must hold both on- and off-shell:
\begin{equation}
    \omega_\lambda = \text{d}\text{k}_\lambda \, .
\end{equation}
Combined with the ambiguity condition~\eqref{eq. cornerambi}, this shows that such charges are naturally identified with corner terms in the symplectic structure. The same reasoning can be applied to any generator not tied to a genuine dynamical constraint.

Another way to phrase it is the following. The symplectic form evaluated along gauge parameters must be conserved on-shell, $\text{d}\omega_\lambda \approx 0$. For unconstrained parameters, this conservation necessarily extends off-shell as well,
\begin{equation}
    \text{d}\omega_\lambda = 0 \, .
\end{equation}
By Poincar\'e lemma, the corresponding symplectic form is then exact and can only contribute through a corner ambiguity~\eqref{eq. cornerambi}.

Finally, since in our case these kinematical fields transform under the residual symmetries~\eqref{eq. s3residualsymm} as independent boundary functions, the corresponding corner ambiguities~$\text{C}$ in the presymplectic potential~\eqref{eq. IWambi} can be directly identified, exactly as implemented in~\eqref{eq. Thetartot}.

%%%%%%%%%%%%%%%%%%%%%%%%%%%%%%%%%%%%%%

\section{Flat Limit} \label{sec. FL}

In this last Section, we construct a smooth flat limit of the asymptotically AdS analysis developed above. Our motivation is twofold. First, the boundary conditions we imposed encompasses the covariant Bondi ones for pure gravity.\footnote{See the related discussion below Eq.~\eqref{eq. RadialGaugeFixing2}.} Second, in this setting a fluid/gravity interpretation~\cite{Campoleoni:2018ltl,Ciambelli:2020eba,Campoleoni:2023fug} prescribes how the fields behave for small values of the boundary light velocity, $k = \tfrac{1}{\ell}$. Here we extend this prescription to higher-spin couplings, thereby providing a first step towards an asymptotically flat description of such spin three anomalous interactions.

Concretely, we prescribe the $k \to 0$ behavior of the field content 
$\psi_\pm = \{\ell_\pm,w_\pm,K^{L,W}_\pm,\phi^{L,W}_\pm\}$ as
\begin{subequations}
    \begin{align}
        &M_L(\theta) = \lim_{k\to0} \left( \ell_+ + \ell_- \right) ,\\
        &N_L(\theta) - t \,\partial_\theta M_L(\theta) = - \lim_{k\to0} \left( \frac{\ell_+ - \ell_-}{k} \right) ,\\
        &M_W(\theta) = \lim_{k\to0} \left( w_+ + w_- \right) ,\\
        &N_W(\theta) - t \,\partial_\theta M_W(\theta) = - \lim_{k\to0} \left( \frac{w_+ - w_-}{k} \right) ,\\
        &\hat{\phi}_\pm^{L,W}(t,\theta) = \lim_{k\to0} \left( \frac{1}{k} \phi_\pm^{L,W} \right) ,\\
        &\hat{K}_\pm^{L,W}(t,\theta) = \lim_{k\to0} \left( \frac{1}{k} K_\pm^{L,W} \right) .
    \end{align}
\end{subequations}
In this way, the flat-space counterparts of the DS generators are identified with the Bondi mass and angular momentum aspects, $M_L$ and $N_L$, together with their higher-spin analogues $M_W$ and $N_W$~\cite{Campoleoni:2017mbt,Campoleoni:2017qot,Campoleoni:2020ejn,Campoleoni:2025bhn}. In the same vein, for the residual symmetry parameters $\xi_\pm = \{Y^\pm,\chi^\pm,H_\pm^{L,W},\omega_\pm^{L,W}\}$, we impose
\begin{subequations}
    \begin{align}
        &Y^\pm = Y_L(\theta) \pm k \left( T_L(\theta) + t \partial_\theta Y_L(\theta) \right) ,\\
        &\chi^\pm = Y_W(\theta) \pm k \left( T_W(\theta) + t \partial_\theta Y_W(\theta) \right) ,\\
        &\hat{\sigma}_\pm^{L,W}(t,\theta) = \lim_{k\to0} \left( \frac{1}{k} \sigma_\pm^{L,W} \right) ,\\
        &\hat{H}_\pm^{L,W}(t,\theta) = \lim_{k\to0} \left( \frac{1}{k} H_\pm^{L,W} \right) ,
    \end{align}
\end{subequations}
where $T_L$ and $Y_L$ denote the usual supertranslation and superrotation parameters, and $T_W$ and $Y_W$ are their $s=3$ generalizations.

These prescriptions define, by construction, the asymptotically flat analogue of the boundary conditions~\eqref{eq. s3solspace1}-\eqref{eq. s3solspace2}. In particular, they yield a smooth flat limit of the charge~\eqref{eq. s3asymptoticcharge}:
\begin{align*}
    &\hat{\text{Q}}_{{\Lambda}} = \lim_{k\to0} \text{Q}_\Lambda \stepcounter{equation}\tag{\theequation} \label{eq. flatcharge}\\
    &= \frac{1}{8 \pi G} \int \text{d}\theta \, \Big[ T_L M_L - Y_L N_L + T_W M_W - Y_W N_W\\
    & + \frac{1}{2} \hat{H}_t^L (\hat{\phi}_+^L + \hat{\phi}_-^L) - \frac{1}{2} \hat{K}_t^L (\hat{\sigma}_+^L + \hat{\sigma}_-^L) + \frac{2}{3} \hat{H}_t^W (\hat{\phi}_+^W + \hat{\phi}_-^W)\\
    & - \frac{2}{3} \hat{K}_t^W (\hat{\sigma}_+^W + \hat{\sigma}_-^W) \Big] \, .
\end{align*}
One can see that the first line of Eq.~\eqref{eq. flatcharge} generates the spin three extension of the BMS$_3$ algebra, previously identified in~\cite{Afshar:2013vka,Gonzalez:2013oaa} and which can be obtained as the Carrollian Inönü--Wigner contraction of two copies of the $\mathcal{W}_3$ algebra~\cite{Campoleoni:2016vsh}. Indeed, expanding the symmetry parameters in Fourier modes, $T_{L,W} \sim e^{\text{i}p\theta}$ and $Y_{L,W} \sim e^{\text{i}p\theta}$, one finds
\begin{subequations} \label{eq. flatW3algebra}
    \begin{align}
        &\text{i} \{ \hat{\text{Q}}_{\Lambda^{Y_L}_{p}} , \hat{\text{Q}}_{\Lambda^{Y_L}_{q}} \} = (p-q) \hat{\text{Q}}_{\Lambda^{Y_L}_{p+q}} \, ,\\
        &\text{i} \{ \hat{\text{Q}}_{\Lambda^{Y_L}_{p}} , \hat{\text{Q}}_{\Lambda^{T_L}_{q}} \} = (p-q) \hat{\text{Q}}_{\Lambda^{T_L}_{p+q}} + \frac{c_M}{12} p^3 \delta_{p+q,0} \, ,\\
        &\text{i} \{ \hat{\text{Q}}_{\Lambda^{Y_L}_{p}} , \hat{\text{Q}}_{\Lambda^{Y_W}_{q}} \} = (2p-q) \hat{\text{Q}}_{\Lambda^{Y_W}_{p+q}} \, ,\\
        &\text{i} \{ \hat{\text{Q}}_{\Lambda^{Y_L}_{p}} , \hat{\text{Q}}_{\Lambda^{T_W}_{q}} \} = (2p-q) \hat{\text{Q}}_{\Lambda^{T_W}_{p+q}} \, ,\\
        &\text{i} \{ \hat{\text{Q}}_{\Lambda^{T_L}_{p}} , \hat{\text{Q}}_{\Lambda^{T_W}_{q}} \} = (2p-q) \hat{\text{Q}}_{\Lambda^{T_W}_{p+q}} \, ,\\
        \begin{split}
        &\text{i} \{ \hat{\text{Q}}_{\Lambda^{Y_W}_{p}} , \hat{\text{Q}}_{\Lambda^{Y_W}_{q}} \} = (p-q) (2p^2 + 2q^2 - pq) \hat{\text{Q}}_{\Lambda^{Y_L}_{p+q}}\\
        &+ \frac{192}{c_M} (p-q) \hat{\text{H}}_{\Lambda^{T_LY_L}_{p+q}} - \frac{44}{5} \frac{96}{{c_M}^2} (p-q) \hat{\text{H}}_{\Lambda^{T_L}_{p+q}} \, ,
        \end{split}\\
        \begin{split}
        &\text{i} \{ \hat{\text{Q}}_{\Lambda^{Y_W}_{p}} , \hat{\text{Q}}_{\Lambda^{T_W}_{q}} \} = (p-q) (2p^2 + 2q^2 - pq) \hat{\text{Q}}_{\Lambda^{T_L}_{p+q}}\\
        &+ \frac{96}{c_M} (p-q) \hat{\text{H}}_{\Lambda^{T_L}_{p+q}} + \frac{c_M}{12} p^5 \delta_{p+q,0} \, ,
        \end{split}
    \end{align}
\end{subequations}
where the central extension is given by $c_M = \tfrac{3}{G}$. We have also used the shorthands $\hat{\text{H}}_{\Lambda^{T_L}_{p}} = \sum_{q\in\mathbb{Z}} \hat{\text{Q}}_{\Lambda^{T_L}_{q}} \hat{\text{Q}}_{\Lambda^{T_L}_{p-q}}$ and  $\hat{\text{H}}_{\Lambda^{T_LY_L}_{p}} = \sum_{q\in\mathbb{Z}}:~\hat{\text{Q}}_{\Lambda^{Y_L}_{q}} \hat{\text{Q}}_{\Lambda^{T_L}_{p-q}}~: - \tfrac{3}{10} (p+2)(p+3) \hat{\text{Q}}_{\Lambda^{T_L}_{p}}$, with normal ordering defined as $:\hat{\text{Q}}_{\Lambda^{Y_L}_{p}} \hat{\text{Q}}_{\Lambda^{T_L}_{q}}: = \hat{\text{Q}}_{\Lambda^{Y_L}_{p}} \hat{\text{Q}}_{\Lambda^{T_L}_{q}}$ for $p<-1$, and with the reversed ordering otherwise.

The remaining non-conserved charges in Eq.~\eqref{eq. flatcharge} generate the flat-space counterparts of the anomalous algebra~\eqref{eq. anomalalgebra}:
\begin{subequations} \label{eq. flatanomalalgebra}
    \begin{align}
        \text{i} \{\hat{\text{Q}}_{\Lambda^{\hat{\sigma}_\pm^L}_{p,q}}, \hat{\text{Q}}_{\Lambda^{\hat{H}_t^L}_{r,s}}\} &= \frac{c_M}{24} \text{e}^{2 \text{i} (q+s) t} \delta_{p+r,q+s} \, ,\\
        \text{i} \{\text{Q}_{\Lambda^{\hat{\sigma}_\pm^W}_{p,q}}, \text{Q}_{\Lambda^{\hat{H}_t^W}_{r,s}}\} &= \frac{c_M}{18} \text{e}^{2 \text{i} (q+s) t} \delta_{p+r,q+s} \, ,
    \end{align}
\end{subequations}
where the arbitrary boundary gauge parameters are expanded in Fourier modes as $\hat{\sigma}_\pm^{L,W} \sim e^{\text{i}(p-q)\theta} e^{\text{i}(p+q)t}$ and  $\hat{H}_t^{L,W} \sim e^{\text{i}(p-q)\theta} e^{\text{i}(p+q)t}$. Only the Heisenberg subalgebras survive in the flat limit, again appearing as one-parameter families of algebras. The putative boundary field theory is expected to be realized in these representations \eqref{eq. flatW3algebra}-\eqref{eq. flatanomalalgebra}. Moreover, it discloses an anomaly: the on-shell variation of the total action~\eqref{eq. Thetartot} admits a non-vanishing smooth flat limit,
\begin{equation} \label{eq. flatanomaly}
    \delta S_\mathrm{tot} \approx - \frac{1}{16 \pi G} \int \text{d}^2x \, \partial_t \left( \hat{K}_t^L \delta (\hat{\phi}_+^L + \hat{\phi}_-^L) + \frac{4}{3} \hat{K}_t^W \delta (\hat{\phi}_+^W + \hat{\phi}_-^W) \right) .
\end{equation}
Since this contribution can once again be interpreted as a corner ambiguity in the associated symplectic structure, the discussion of Subsec.~\ref{subsec. ST} applies equally well in the present flat-space context. Notably, only the components \(\hat{K}_t^{L,W}\) play an anomalous role in the flat limit.

%%%%%%%%%%%%%%%%%%%%%%%%%%%%%%%%%%%%%%

\section{Conclusion} \label{sec. Conclu}

We have identified a new class of boundary conditions~\eqref{eq. s3solspace1}-\eqref{eq. s3solspace2} for the Chern--Simons formulation of three-dimensional asymptotically Anti--de Sitter gravity coupled to massless higher-spin fields. These conditions rely on arbitrary kinematical boundary fields and thus generalize earlier constructions in three-dimensional Einstein--Hilbert gravity~\cite{Compere:2013bya,Troessaert:2013fma,Perez:2016vqo,Grumiller:2016pqb,Campoleoni:2018ltl,Alessio:2020ioh,Geiller:2021vpg,Campoleoni:2022wmf,Ciambelli:2023ott,Delfante:2024npo}. It enhances the Zamolodchikov algebra of charges~\eqref{eq. W3algebra}-\eqref{eq. anomalalgebra} via centrally extended Abelian sectors, enlarging the space of physical symmetries and offering stronger algebraic tools to organize the observables of the theory.

The analysis was carried out in the radial gauge~\eqref{eq. RadialGaugeFixing}, where both the on-shell action and the surface charges are automatically finite at asymptotic infinity, without requiring holographic renormalization. This property may be viewed as a CS analogue of Gauss’s theorem: the surface charge associated with a given source can be computed at any radial distance, provided no new sources are crossed. It would be interesting to connect this observation with recent progress on coadjoint orbits of maximal corner charges~\cite{Ciambelli:2021vnn,Ciambelli:2022cfr} and symplectic prescriptions~\cite{Freidel:2019ohg,McNees:2023tus,Campoleoni:2023eqp,Riello:2024uvs,Campoleoni:2025bhn}. Although the radial gauge can always be imposed off shell near the boundary---and is argued to introduce no new physics in standard settings~\cite{Grumiller:2016pqb,Ciambelli:2023ott}---its relaxation may reveal additional boundary structures.

In fact, fields generated by such gauge relaxations typically may remain unconstrained by the equations of motion and thus behave as kinematical charges. As shown in Subsec.~\ref{subsec. ST}, such contributions can be added or removed from the total asymptotic charge through corner redefinitions of the Iyer--Wald symplectic structure~\cite{Lee:1990nz,Wald:1993nt,Iyer:1994ys}. What is appealing is that the anomalous charges~\eqref{eq. s3asymptoticcharge} constructed in this work are also precisely of this type. Besides, they induce anomalies in the variational principle~\eqref{eq. Thetartot}. Therefore, this raises the broader question of which corner ambiguities should be considered physically meaningful, a problem that recurs across higher dimensions and in a variety of gauge theories. Progress here requires a refined understanding of boundary geometry and its holographic dual, as emphasized in~\cite{Capone:2023roc,Riello:2024uvs,Campoleoni:2025bhn}. Our results provide preliminary evidence that anomalous boundary fields in higher-spin gravity connect directly to gauge relaxations already studied in pure gravity, such as the Weyl--Fefferman--Graham~\cite{Ciambelli:2019bzz,Ciambelli:2023ott} and covariant Bondi~\cite{Campoleoni:2018ltl,Ciambelli:2020eba,Campoleoni:2022wmf,Campoleoni:2023fug} gauges.

A systematic treatment of these issues will likely require embedding them in the Toda framework that governs higher-spin couplings in $\mathcal{W}$-holography. For instance, the Miura transformation~\eqref{eq. Miura}, central to our generalization of boundary conditions from $s=2$ to $s=3$, originates in this formalism, where the quantities $\vec{\mathcal{Q}}$ coincide with the Toda charges of the dual model. This perspective echoes earlier analyses of Weyl anomalies in higher-spin systems~\cite{Li:2015osa} and may prove essential for interpreting holographic duals. It would also be interesting to examine whether our extended solution space~\eqref{eq. s3solspace1}-\eqref{eq. s3solspace2} can support higher-spin black hole solutions \cite{Gutperle:2011kf,Bunster:2014mua}, for instance by radially reconstructing the generalized Fefferman–Graham gauge \cite{Grumiller:2016pqb,Krishnan:2017xct,Arenas-Henriquez:2024ypo}, which is built from a different group element than the one used in \ref{sec. RC} and specifically designed to incorporate chemical potentials.

Finally, we have outlined a smooth flat limit of our higher-spin analysis, which yields both the Heisenberg enhancement of the spin-three BMS algebra \eqref{eq. flatanomalalgebra} and the prediction of an associated anomaly \eqref{eq. flatanomaly} in the putative dual theory. A natural next step is to extend the fluid/gravity correspondence in the covariant Bondi gauge~\cite{Campoleoni:2018ltl,Ciambelli:2020eba,Campoleoni:2022wmf,Campoleoni:2023fug} to higher-spin theories. In this framework, the bulk flat limit maps the boundary theory from a relativistic to a Carrollian hydrodynamic system. Understanding how higher-spin couplings deform this transition could sharpen the structure of the dual field theory and, at the same time, yield fresh insights into fluid dynamics.\footnote{See, for instance, \cite{Lee:2013gwa} 
for a conjecture on the role of higher spin couplings in the dissipative gradient of the stress tensor of the dual boundary fluid.}

%%%%%%%%%%%%%%%%%%%%%%%%%%%%%%%%%%%%%%

\section*{Acknowledgments}

The author would like to thank Andrea Campoleoni, Luca Ciambelli and Chrysoula Markou for useful discussions and  and feedbacks on the manuscript. His work was supported by the Belgian American Educational Foundation.

%%%%%%%%%%%%%%%%%%%%%%%%%%%%%%%%%%%%%%

\appendix

\section{Radial Reconstruction} \label{sec. RC}

In this Appendix we reconstruct the radial coordinate so that it acquires the same geometric interpretation as in the two canonical gauges of three-dimensional gravity: the Fefferman--Graham gauge~\cite{FG1} and the Bondi gauge~\cite{Bondi:1960jsa,Sachs:1961zz}. We then extend the analysis to higher-spin couplings. This procedure highlights the analogous gauge relaxations induced by our new asymptotic solution space~\eqref{eq. s3solspace1}-\eqref{eq. s3solspace2} in the second-order formalism.

\subsection{Spacelike Coordinate}

We consider the gauge transformations~\eqref{eq. RadialGaugeFixing} acting on the radially independent solution space~\eqref{eq. s3solspace1}-\eqref{eq. s3solspace2}, implemented through the $SL(3,\mathbb{R})$ group elements
\begin{equation} \label{eq. FG radial}
    b(z) = \exp \left( - \ell \log z \, L_0 \right) , \quad \bar{b}(z) = b^{-1}(z) \, .
\end{equation}
For $s=2$, the metric follows from the CS connections as
\begin{equation} \label{eq. gmunu}
    g_{\mu\nu} = 2 \, \mathrm{tr} \left( e_\mu e_\nu \right) ,
\end{equation}
where the bulk dreibein is defined by
\begin{equation}
    \text{e} = \frac{\ell}{2} \left( \text{A} - \bar{\text{A}} \right) \, .
\end{equation}
With the choice~\eqref{eq. FG radial}, the radial reconstruction yields
\begin{equation}
    \text{d}s^2 = \frac{\ell^2}{z^2} \text{d}z^2 + \frac{2}{z} g_{za}(x^b) \text{d}z \, \text{d}x^a  + \frac{1}{z^2} g_{ab}(z,x^c) \text{d}x^a \text{d}x^b \, ,
\end{equation}
so that $z$ is a spacelike radial coordinate and $x^a = (t,\theta)$ are the boundary coordinates.

A key feature of this construction is that the usual FG condition $g_{za}=0$ is relaxed, leading to
\begin{equation} \label{eq. gzpm}
    g_{z\pm} = - \ell^2 \left( K_\pm^L + \frac{1}{2} \partial_\pm \left( \phi_+^L + \phi_-^L \right) \right) .
\end{equation}
In the WFG restriction, 
$\{\phi_\pm^L = \varphi, \; K_\pm^L = k_\pm - \partial_\pm \varphi\}$, this simplifies to $g_{z\pm} = k_\pm$. The relaxation is thus entirely governed by the Weyl connection, which restores Weyl covariance at the boundary~\cite{Ciambelli:2019bzz,Ciambelli:2023ott}. In contrast, the standard FG analysis~\cite{Imbimbo:1999bj,Rooman:2000zi} explicitly breaks this symmetry. Here, instead, the boundary acquires a full Weyl geometry, with a dynamical boundary metric. In this reduced setting, the boundary metric typically appears at leading order in the asymptotic expansion,
\begin{equation} \label{eq. gab}
    g_{ab} = g_{ab}^{(0)}(x^c) + z^2 g_{ab}^{(2)}(x^c) + z^4 g_{ab}^{(4)}(x^c) \, ,
\end{equation}
with successive coefficients determined by Eq.~\eqref{eq. gmunu}. The first subleading term encodes the boundary one-point function via the stress tensor, while the last is locally fixed in terms of the first two. See~\cite{Banados:2004nr,Ciambelli:2023ott} for further details in (Weyl--)Fefferman--Graham gauges.

These considerations extend naturally to the spin-3 metric-like field \cite{Campoleoni:2010zq,Campoleoni:2012hp,Fredenhagen:2014oua,Campoleoni:2014tfa},
\begin{equation} \label{eq. varphimunurho}
    \varphi_{\mu\nu\rho} = \frac{2}{3} \text{tr} \left( e_\mu e_\nu e_\rho \right) \, ,
\end{equation}
which admits the asymptotic form
\begin{align*}
    &\varphi_{\mu\nu\rho} \text{d}x^\mu \text{d}x^\nu \text{d}x^\rho = \stepcounter{equation}\tag{\theequation}\\
    &\frac{1}{z^2} \left( \varphi_{zza}(x^b) \text{d}z^2 + \frac{\text{d}z}{z} \varphi_{zab}(z,x^c) \text{d}x^b + \varphi_{abc}(z,x^d) \text{d}x^b \text{d}x^c \right) \text{d}x^a \, .
\end{align*}
Compared to the radial gauge in Poincar\'e coordinates (see, e.g.,~\cite{Campoleoni:2016uwr}), one finds the relaxed condition
\begin{equation}
    \varphi_{zz\pm} = \frac{\ell^3}{6} \left( 2 K_\pm^W + \partial_\pm \left( \phi_+^W + \phi_-^W \right) \right) ,
\end{equation}
while the other components exhibit finite radial expansions,
\begin{equation}
    \varphi_{zab} = \sum_{n  = 0}^2 z^{2n} \varphi_{zab}^{(2n)}(x^c) \, , \qquad \varphi_{abc} = \sum_{n = 0}^3 z^{2n} \varphi_{abc}^{(2n)}(x^d) \, .
\end{equation}

\subsection{Null Coordinate}

A parallel construction can be carried out for a null radial coordinate $r$, with the conformal boundary located at $r \to \infty$. The reconstruction is implemented via~\eqref{eq. RadialGaugeFixing}, supplemented by the $SL(3,\mathbb{R})$ group elements
\begin{equation}
    b(r) = \exp \left( r \, k \, L_{-1} \right) , \quad \bar{b}(r) = \mathbb{I} \, ,
\end{equation}
with $k = \tfrac{1}{\ell}$. This choice produces a Bondi-like line element,
\begin{equation}
    \text{d}s^2 = 2 g_{ra}(x^b) \text{d}r \text{d}x^a + r^2 g_{ab}(r,x^c) \text{d}x^a \text{d}x^b \, ,
\end{equation}
where the standard Bondi condition $g_{r\theta}=0$ is relaxed to
\begin{equation}
    g_{r\theta} = \frac{1}{2k} \left( \text{e}^{\phi_+^L} \cosh\left(2\phi_+^W\right) + \text{e}^{\phi_-^L} \cosh\left(2\phi_-^W\right) \right) .
\end{equation}
In the restriction $\{\phi^L_\pm = \varphi \pm \zeta, \; K_\pm^L = \mp \partial_\pm \zeta\}$, this relation identifies with the transverse fluid velocity, thereby distinguishing hydrodynamic frames~\cite{Ciambelli:2020eba}. In this setting, the fluid velocity also appears at leading order in the finite radial expansion,
\begin{equation}
    g_{ab} = g_{ab}^{(0)}(x^c) + \frac{1}{r} g_{ab}^{(1)}(x^c) + \frac{1}{r^2} g_{ab}^{(2)}(x^c) \, ,
\end{equation}
so that the first subleading term contributes to the Weyl connection, while the last encodes the boundary fluid stress tensor. For further details on (covariant) Bondi gauges, see~\cite{Campoleoni:2018ltl,Ciambelli:2020eba,Campoleoni:2022wmf,Campoleoni:2023fug}.

The corresponding spin-3 metric-like field then takes the form
\begin{align*}
    &\varphi_{\mu\nu\rho} \text{d}x^\mu \text{d}x^\nu \text{d}x^\rho = \stepcounter{equation}\tag{\theequation}\\
    &r \left( \varphi_{rab}(r,x^c) \text{d}r + r^2 \varphi_{abc}(r,x^d) \text{d}x^c \right) \text{d}x^a \text{d}x^b \, ,
\end{align*}
with finite radial expansions following Eq.~\eqref{eq. varphimunurho}:
\begin{equation}
    \varphi_{rab} = \sum_{n = 0}^1 \frac{1}{r^n}\varphi_{rab}^{(n)}(x^c) \, , \qquad \varphi_{abc} = \sum_{n = 0}^3 \frac{1}{r^n}\varphi_{abc}^{(n)}(x^d) \, .
\end{equation}
This structure is likewise gauge-relaxed with respect to the higher-spin Bondi-like gauge~\cite{Campoleoni:2017mbt,Campoleoni:2017qot,Campoleoni:2020ejn,Campoleoni:2025bhn}. In particular, one finds the non-vanishing component
\begin{equation}
    \varphi_{r\theta\theta} = - \frac{r}{2k} \text{e}^{\phi_+^L+\phi_-^L} \sinh\left( 2 \left( \phi_+^W - \phi_-^W \right) \right) + \mathcal{O}(1) \, .
\end{equation}
%

%%%%%%%%%%%%%%%%%%%%%%%%%%%%%%%%%%%%%%

\providecommand{\href}[2]{#2}\begingroup\raggedright\endgroup

\end{document}